\documentclass[sigconf]{acmart}

\AtBeginDocument{%
  }
\usepackage{adjustbox}
\usepackage{bbm}
\usepackage{enumitem}
\usepackage{balance}
\usepackage{hyperref}

\copyrightyear{2025}
\acmYear{2025}
\setcopyright{cc}
\setcctype{by}
\acmConference[SIGIR '25]{Proceedings of the 48th International ACM SIGIR
Conference on Research and Development in Information Retrieval}{July 13--18,
2025}{Padua, Italy}
\acmBooktitle{Proceedings of the 48th International ACM SIGIR Conference on
Research and Development in Information Retrieval (SIGIR '25), July 13--18,
2025, Padua, Italy}\acmDOI{10.1145/3726302.3730192}
\acmISBN{979-8-4007-1592-1/2025/07}

\begin{document}

\title{Exploring $\ell_0$ Sparsification for Inference-free Sparse Retrievers}

\author{Xinjie Shen}
\orcid{0009-0004-9176-5400}
\authornote{Work done during internship at Amazon. Corresponding author: Yang Yang.}
\affiliation{%
  \institution{South China University of Technology}
  \department{School of Future Technology}
  \city{Guangzhou}
  \country{China}}
\affiliation{%
  \institution{Amazon}
  \department{Amazon Web Service}
  \city{Shanghai}
  \country{China}}
\email{202164690138@mail.scut.edu.cn}

\author{Zhichao Geng}
\orcid{0009-0005-8100-2079}
\affiliation{%
  \institution{Amazon}
  \department{Amazon Web Service}
  \city{Shanghai}
  \country{China}}
\email{zhichaog@amazon.com}

\author{Yang Yang}
\orcid{0009-0005-2480-3339}
\affiliation{%
  \institution{Amazon}
  \department{Amazon Web Service}
  \city{Shanghai}
  \country{China}}
\email{yych@amazon.com}

\renewcommand{\shortauthors}{Shen et al.}

\begin{abstract}
With increasing demands for efficiency, information retrieval has developed a branch of sparse retrieval, further advancing towards inference-free retrieval where the documents are encoded during indexing time and there is no model-inference for queries. Existing sparse retrieval models rely on FLOPS regularization for sparsification, while this mechanism was originally designed for Siamese encoders, it is considered to be suboptimal in inference-free scenarios which is asymmetric. Previous attempts to adapt FLOPS for inference-free scenarios have been limited to rule-based methods, leaving the potential of sparsification approaches for inference-free retrieval models largely unexplored. In this paper, we explore $\ell_0$ inspired sparsification manner for inference-free retrievers. Through comprehensive out-of-domain evaluation on the BEIR benchmark, our method achieves state-of-the-art performance among inference-free sparse retrieval models and is comparable to leading Siamese sparse retrieval models. Furthermore, we provide insights into the trade-off between retrieval effectiveness and computational efficiency, demonstrating practical value for real-world applications.
\end{abstract}



\received{20 February 2007}
\received[revised]{12 March 2009}
\received[accepted]{5 June 2009}

\maketitle

\section{Introduction}

Information retrieval systems have evolved significantly over the past decades, with sparse retrieval remaining a cornerstone branch due to its efficiency, scalability, and reasonable search relevance\cite{effcientSplade}. While dense retrievers such as ColBERTv2\cite{santhanam-etal-2022-colbertv2} and Contriever\cite{izacard2021contriever} learn continuous semantic representations of queries and documents through neural networks, sparse retrievers focus on capturing lexical matches through token-based representations\cite{10.1145/3397271.3401204,MacAvaney2020,zhao2021sparta}. 

To further enhance retrieval efficiency, inference-free retrieval methods\cite{zhichao,splade3, nogueira2019document,dai2020context} have emerged as a promising solution to the computational challenges in modern search systems\cite{Banon2025opensearch}. These approaches pre-compute document representations while degenerating query-time inference into lightweight methods such as term matching\cite{https://doi.org/10.1002/asi.4630270302,10.1561/1500000019} or tokenization\cite{splade1,splade2,splade2bis,splade3}. This design achieves superior efficiency while maintaining competitive performance.

The efficiency of sparse retrieval models is largely determined by their sparsification capability. A key design element in Siamese sparse retriever models is the FLOPS (Floating Point Operations) regularization\cite{Paria2020Minimizing}, which penalizes non-zero elements in sparse representations using their squared average. However, this approach is less suited for inference-free models due to their asymmetric architecture, which differs from the original Siamese design in models like SPLADE. Moreover, FLOPS continually penalizes the scale of token weights but pays less attention to tokens' $\ell_0$. The encoded document length could lower to zero with larger weight on FLOPS,

In this paper, we propose a combination of two novel approaches designed for sparse inference-free models: $\ell_0$ mask FLOPS and $\ell_0$ approximation activation. These techniques introduce selective sparsification by applying regularization only to document side representations exceeding desired sparsity thresholds, while allowing already-sparse representations to optimize unrestrictedly for the ranking objective. Through extensive out-of-domain experiments on BEIR, we demonstrate that we can achieve superior performance to representative dense retriever models and inference-free models, and become comparable with state-of-the-art sparse retriever models, while maintaining good computational efficiency.

The key contributions of our work include:
\begin{itemize}[noitemsep,topsep=0pt,leftmargin=2em]
\item Two novel sparsification techniques, $\ell_0$ mask loss and $\ell_0$ approximation activation, designed for inference-free retrievers.
\item Strong empirical results on BEIR benchmark, achieving comparable performance to state-of-the-art sparse retrievers.
\item Comprehensive analysis of efficiency-performance trade-offs, providing practical insights for real-world scenarios.
\end{itemize}
\section{Related Work}
\subsection{Sparse Retrievers}
Compared to dense retrievers, like ColBERTv2\cite{santhanam-etal-2022-colbertv2}, Contriever\cite{izacard2021contriever} and TAS-B\cite{TAS-B}, which learn continuous semantic representations of both queries and documents utilizing neural networks, sparse representations focus on capture lexical matches through discrete token-based representations\cite{10.1145/3397271.3401204,https://doi.org/10.48550/arxiv.2010.00768,MacAvaney2020,zhao2021sparta}. Specifically, sparse representations typically represent documents and queries as high-dimensional sparse vectors where each dimension corresponds to a specific term in the vocabulary, and the value indicates the importance of that term. The SPLADE series models\cite{splade1,splade2,splade2bis,splade3} have achieved great success as representative sparse retriever models, utilizing BERT's masked language model head and pooling to generate vocabulary-sized sparse representations. However, query encoding remains an efficiency bottleneck in online practice scenarios, requiring further development and variants for adaptation.

\subsection{Inference-free Retrievers}
Inference-free retriever models, such as BM25\cite{https://doi.org/10.1002/asi.4630270302,10.1561/1500000019}, DEEPCT\cite{dai2020context}, Doc2Query\cite{nogueira2019document}, DeepImpact\cite{deepimpact}, EPIC\cite{EPIC}, SPLADE-v3-Doc\cite{splade3} and SPLADE-doc-distill\cite{zhichao,splade2}, can directly retrieve relevant documents without requiring computationally expensive neural inference on the query side. These models typically rely on pre-computed document representations offline, and simple operations for obtaining query representations online, making them particularly attractive for real-world applications\cite{6228205, 10.5555/3235358} where latency and computational resources are critical concerns due to huge query volumes. Unlike inference-required models that need to encode queries through neural networks during retrieval, inference-free retrievers use simple operations like term matching\cite{https://doi.org/10.1002/asi.4630270302,10.1561/1500000019} or tokenization\cite{splade1,splade2,splade2bis,splade3} to avoid huge computational costs, while offering lower but reasonable throughput and latency.

\section{Preliminary}
Our work is built upon the representative sparse inference-free retriever, \textit{SPLADE-doc-distill}. This model predicts token importance across the vocabulary space using BERT's masked language model head. The model creates sparse 30,522-dimensional vectors using max pooling and ReLU activation. It operates asymmetrically, processing only documents. Moreover, we adopt the IDF-aware technique proposed by \cite{zhichao}. For an input document with token weight ($w_{i,j}$) at position ($i$), the sparse representation of token $j$ is:
\begin{equation}
     w_j = \text{IDF}_j \cdot \max_i \log(1 + \sigma(w_{i,j})),
\label{eq:1}
\end{equation}
where $\text{IDF}_j$ is the IDF value of token $j$ and $\sigma(\cdot)$ is the ReLU function.

For query representation, it employs an inference-free approach to generate a Bag-of-Words-like sparse vector, where the weight \(w_j\) of token \(j\) in the query representation is set to 1 only if token \(j\) is present in the query; otherwise, \(w_j\) is set to 0. IDF values are then applied for adjustment as well. The ranking score for a given query and document can be computed as the inner product of their sparse representations. The optimization objective of SPLADE combines the ranking loss \(\mathcal{L}_{\text{rank}}\) and sparse regularization loss \( \mathcal{L}^{d}_{\texttt{FLOPS}}\) for optimizing search relevance and representation sparsification. For \textit{IDF-SPLADE-doc-distill}, the training loss \(\mathcal{L}\) is formulated as follows:
\begin{equation}
\mathcal{L} = \mathcal{L}_{\texttt{rank}} + \lambda_d \mathcal{L}^{d}_{\texttt{FLOPS}}, \\
\label{eq:add_loss}
\end{equation}
\begin{equation}
\mathcal{L}^{d}_{\texttt{FLOPS}} =\sum_{j\in V} {\bar a}_j^2 = \sum_{j \in V} \left( \frac{1}{N} \sum_{i=1}^N  \frac{w_j^{(d_i)}}{\text{IDF}_j} \right)^2,
\label{eq:loss}
\end{equation}
where \(\lambda_d\) is the weight for the FLOPS regularizer, $d_i$ is the $i$-th document in batch and \(\overline{a}_j\) is the average weight of token \(j\) in the batch. The FLOPS regularizer penalizes dimensions with high average weights in the representation to achieve sparsity. We denoted this base model as \textit{IDF-SPLADE-doc-distill} in the following.

\section{Method}
{\let\thefootnote\relax\footnotetext{Code is available at this Github repository: \url{https://github.com/zhichao-aws/opensearch-sparse-model-tuning-sample/tree/l0_enhance}. All details are included.}}

\subsection{$\ell_0$ Mask Loss}
\label{l0mask}

Given the loss format in Equation \ref{eq:add_loss} for inference-free sparse retriever, it can be observed that with a relatively large weight of $\lambda_d$, the sparsification objective may easily dominate the $\mathcal{L}_{\texttt{rank}}$ objective, pushing the average encoded sparse representations' $\ell_0$ to a very limited amount that harms the learning of $\mathcal{L}_{\texttt{rank}}$. In this section, we propose a threshold method named $\ell_0$ Mask Loss, which focuses on utilizing the advantages of FLOPS loss in achieving sparsity while avoiding further reduction of token weights and the $\ell_0$ of sparse representations (e.g., collapsing to 0), which can harm the learning of the ranking objective, once reasonable sparsity is achieved.

Specifically, we calculate the $\ell_0$ of each $w^{(d_i)}$ in the batch and build a binary mask based on a given threshold $t$. If the amount of activated tokens in $w^{(d_i)}$ is already lower than the given threshold, then $w^{(d_i)}$ will not participate in the calculation of FLOPS loss, allowing free learning without the penalty and constraints on token weight. The $\mathcal{L}^{d}_{\texttt{FLOPS}}$ can be rewritten as follows:
\begin{align}
{\mathcal{L}^{d}_{\texttt{FLOPS}}}' &= \sum_{j \in V} \left( \frac{1}{N} \sum_{i=1}^N M^{(d_i)} \frac{w^{(d_i)}_{j}}{\text{IDF}_j} \right)^2, \\
M^{(d_i)} &= \mathbbm{1}[\|w^{(d_i)}\|_0 > t],
\end{align}
where $M^{(d_i)}$ is a binary mask indicator vector that equals $\mathbbm{1} \in \mathbb{R}^{|V|}$ if the $\ell_0$ norm of weight vector $w^{(d_i)}$ exceeds threshold $t$, and all zeros otherwise. This mask is designed to effectively exclude already-sparse document side presentations from the FLOPS loss calculation, which helps to provide specific focus for the model under asymmetric structure.

\subsection{$\ell_0$ Approximation Activation}
\label{l0activation}
From Equation ~\ref{eq:loss}, it can also be observed that FLOPS largely relies on punishing representations via their weights' scale that related to the retrieval time\cite{splade2}, rather than focusing on the $\ell_0$ of the representation. This leads to more focus on larger weights and potentially overlooks small weights, since the resulting gradients have diverse differences. To help the regularizer focus more on small weights rather than being overwhelmingly led by large gradients of large weights, we suggest using multiple subsequent log transformations in activation. Compared to the previous design, such modification shifts more attention to the $\ell_0$ of the representation rather than the token scale, therefore we named it as $\ell_0$ approximation activation here. With applying such modification, larger values grow much slower than a linear function and serve as a soft cap, creating a natural sparsity-inducing effect for the model similar to $\ell_0$ regularization. 

\begin{table*}[h]
\caption{Model performances (NCDG@10) on 13 datasets of BEIR. Best performance of each retriever type are bold.}
\adjustbox{max width=\textwidth}{
\centering
\label{tab:main_eval}
\begin{tabular}{cccccccccccc}
\toprule
               & \multicolumn{6}{c}{\textbf{Inference-free Sparse Retriever}} & \multicolumn{2}{c}{\textbf{Sparse Retriever}} & \multicolumn{3}{c}{\textbf{Dense Retriever}} \\
\cmidrule(lr){2-7} \cmidrule(lr){8-9} \cmidrule(lr){10-12}
Dataset        & $\ell_0$ Mask & \begin{tabular}[c]{@{}c@{}}$\ell_0$ Mask-\\$\ell_0$ Activation\end{tabular}  & \begin{tabular}[c]{@{}c@{}}IDF-SPLADE-\\ doc-distill\end{tabular} & BM25 & \begin{tabular}[c]{@{}c@{}}SPLADE-\\ doc-distill\end{tabular} & \begin{tabular}[c]{@{}c@{}}SPLADE-\\ v3-Doc\end{tabular} & \begin{tabular}[c]{@{}c@{}}SPLADE\\ ++\-SelfDistil\end{tabular} & \begin{tabular}[c]{@{}c@{}}SPLADE\\ -v3-Distil\end{tabular} & ColBERTv2 & Contriever & TAS-B \\
\midrule
TREC-COVID     & 70.4           & 71.3                                                      & 67.0                                                              & 68.8  & 68.4                                                          & 68.1                                                     & 71.0                                                            & 70.0                                                        & \textbf{73.8}    & 59.6            & 48.1    \\
NFCorpus       & 34.1           & 34.1                                                      & 32.6                                                              & 32.7  & 34.0                                                          & 33.8                                                     & 33.4                                                            & \textbf{34.8}                                               & 33.8             & 32.8            & 31.9    \\
NQ             & 54.0           & 54.1                                                      & 52.6                                                              & 32.6  & 48.8                                                          & 52.1                                                     & 52.1                                                            & 54.9                                                        & \textbf{56.2}    & 49.8            & 46.3    \\
HotpotQA       & 67.6           & 68.1                                                      & 67.9                                                              & 60.2  & 62.6                                                          & 66.9                                                     & \textbf{68.4}                                                   & 67.8                                                        & 66.7             & 63.8            & 58.4    \\
FiQA-2018      & 35.4           & 34.6                                                      & 34.2                                                              & 25.4  & 31.2                                                          & 33.6                                                     & 33.6                                                            & 33.9                                                        & \textbf{35.6}    & 32.9            & 30.0    \\
ArguAna        & 48.9           & \textbf{49.7}                                             & 48.6                                                              & 47.2  & 37.7                                                          & 46.7                                                     & 47.9                                                            & 48.4                                                        & 46.3             & 44.6            & 42.9    \\
Touche-2020    & 29.2           & 28.3                                                      & 27.6                                                              & 34.7  & 25.6                                                          & 27.0                                                     & \textbf{36.4}                                                   & 30.1                                                        & 26.3             & 23.0            & 16.2    \\
DBPedia-entity & 42.2           & 42.2                                                      & 41.3                                                              & 28.7  & 35.9                                                          & 36.1                                                     & 43.5                                                            & 42.6                                                        & \textbf{44.6}    & 34.5            & 38.4    \\
SCIDOCS        & 16.1           & 16.3                                                      & 16.4                                                              & 16.5  & 14.7                                                          & 15.2                                                     & 15.8                                                            & 14.8                                                        & 15.4             & \textbf{41.3}   & 14.9    \\
FEVER          & 81.0           & 81.1                                                      & \textbf{81.6}                                                     & 64.9  & 67.4                                                          & 68.9                                                     & 78.6                                                            & 79.6                                                        & 78.5             & 16.5            & 70.0    \\
Climate-FEVER  & 21.9           & 21.1                                                      & 21.5                                                              & 18.6  & 15.1                                                          & 15.9                                                     & 23.5                                                            & 22.8                                                        & 17.6             & \textbf{75.8}   & 22.8    \\
SciFact        & \textbf{71.5}  & 70.6                                                      & 70.5                                                              & 69.0  & 70.8                                                          & 68.8                                                     & 69.3                                                            & 68.5                                                        & 69.3             & 23.7            & 64.3    \\
Quora          & 83.1           & 82.2                                                      & 82.1                                                              & 78.9  & 73.0                                                          & 77.5                                                     & 83.8                                                            & 81.7                                                        & 85.2             & \textbf{86.5}   & 83.5    \\ \hline
Aver.Rank      & 3.54           & \textbf{3.38}                                             & 5.15                                                              & 7.85  & 8.62                                                          & 7.46                                                     & \textbf{3.92}                                                   & 4.69                                                        & \textbf{4.69}    & 7.38            & 8.77    \\
Average        & \textbf{50.43} & 50.28                                                     & 49.52                                                             & 44.48 & 45.02                                                         & 46.97                                                    & \textbf{50.56}                                                  & 49.99                                                       & \textbf{49.95}   & 44.98           & 43.67   \\ \hline
\end{tabular}}
\vspace{-0.2cm}
\end{table*}

Since the original design of the SPLADE model has one log, we suggest rewriting the activation function $\sigma(\cdot)$ in Equation~\ref{eq:1} as follows:
\begin{equation}
\sigma(x) = \log(1 + \max(0,x)).
\end{equation}
We replace the original RELU activation function with our $\ell_0$ approximation activation during both training and evaluation.
\section{Experiment}

In this section, we aim to answer two research questions: \textbf{Q1}: How do our proposed methods improve performance? \textbf{Q2}: How do our methods affect the trade-off between efficiency and performance?
\subsection{Experiment Setup}

\subsubsection{Datasets}
The MS MARCO dataset \cite{msmarco} is employed to finetune\footnote{We use the word finetune since models like SPLADE are initialized from BERT pretrained weights} all models, and we initialize the model from Co-Condenser \cite{gao2022unsupervised} checkpoint. Teacher scores preparation process follows \cite{zhichao}. The MS MARCO dataset comprises 8,841,823 passages and 502,939 queries in the training set. Following the work of \cite{10.1145/3477495.3531857,splade3}, we evaluate our model's \textbf{zero-shot} out-of-domain performance on a readily available subset of 13 datasets from the BEIR benchmark \cite{thakur2021beir}.

\subsubsection{Baselines}
In this section, we include three types of baselines: 1) \textbf{Inference-free sparse retrievers}: BM25\cite{https://doi.org/10.1002/asi.4630270302,10.1561/1500000019}, IDF-SPLADE-doc-distill\cite{zhichao}, SPLADE-doc-distill\cite{zhichao,splade2}, SPLADE-v3-Doc\cite{splade3}, 2) \textbf{Sparse retrievers}: SPLADE++-SelfDistil\cite{splade2bis}, SPLADE-v3-Distil\cite{splade3}, 3) \textbf{Dense retrievers}: ColBERTv2\cite{santhanam-etal-2022-colbertv2}, Contriever\cite{izacard2021contriever}, TAS-B\cite{TAS-B}. During training and evaluation, IDF values are derived from the MS MARCO dataset.

\subsubsection{Metrics and Evaluation}
We include three metrics to measure models' performance and efficiency: 1) NDCG@10, 2) FLOPS\cite{Paria2020Minimizing}, 3) the average number of non-zero tokens in encoded document sparse representation, denoted as Doc\_Len. For the NDCG@10 metric, we calculate it using the BEIR python toolkit.

\begin{table}[h]
\caption{Ablation study in comparable FLOPS.}
\centering
\label{tab:ablation}
\begin{tabular}{cccc}
\hline
Model                  & NCDG@10 & FLOPS & Doc\_Len \\ \hline
Baseline & 49.52   & 2.39  & 327.23   \\
+$\ell_0$ mask                 & 50.43   & 2.31  & 322.22   \\
+$\ell_0$ Activation                &  49.97       &  2.30     &  295.79       \\
$+\ell_0$ mask-$\ell_0$ Activation         & 50.28   & 2.13  & 275.02   \\ \hline
\end{tabular}
\vspace{-0.3cm}
\end{table}

\subsubsection{Indexing}
We use OpenSearch\footnote{\url{https://opensearch.org/}} as our lexical search engine to construct the inverted index and perform the retrieval process. The writing and searching processes for custom learned sparse models are integrated through the OpenSearch neural sparse feature. The maximum input length is set to 512 tokens.

\subsection{(Q1): Relevance Evaluation}

In this section, we present the zero-shot out-of-domain (OOD) evaluation results on the BEIR benchmark, as shown in Table \ref{tab:main_eval}. Our proposed methods demonstrate significant improvement, outperforming the best inference-free sparse retriever. Moreover, our approach outperforms all dense retrievers and maintains performance comparable to state-of-the-art Siamese sparse retrievers, all while retaining inference-free efficiency.

To assess the overall effectiveness across different datasets, we also analyzed the relative ranking of each method. Our approach achieved the best average rank among all three types of retrievers (dense, inference-free sparse, and Siamese sparse), demonstrating its robust performance across diverse domains and retrieval tasks.

From the ablation study provided in Table \ref{tab:ablation}, we can observe that the $\ell_0$ mask remains a good representative with reasonable FLOPS and Doc\_Len while maintaining good performance. Application of $\ell_0$ Activation explicitly improves the efficiency with limited relevance degradation. However, directly applying $\ell_0$ Activation increases efficiency but harms performance, suggesting that $\ell_0$ Activation relies on $\ell_0$ mask. In the following section, we dive deeper into the trade-off between efficiency and performance.

\subsection{(Q2): Efficiency Trade-off Analysis}

\subsubsection{Varying FLOPS Penalty}

\begin{figure}[h]
    \centering
    \vspace{-0.3cm}
    \includegraphics[width=1\columnwidth]{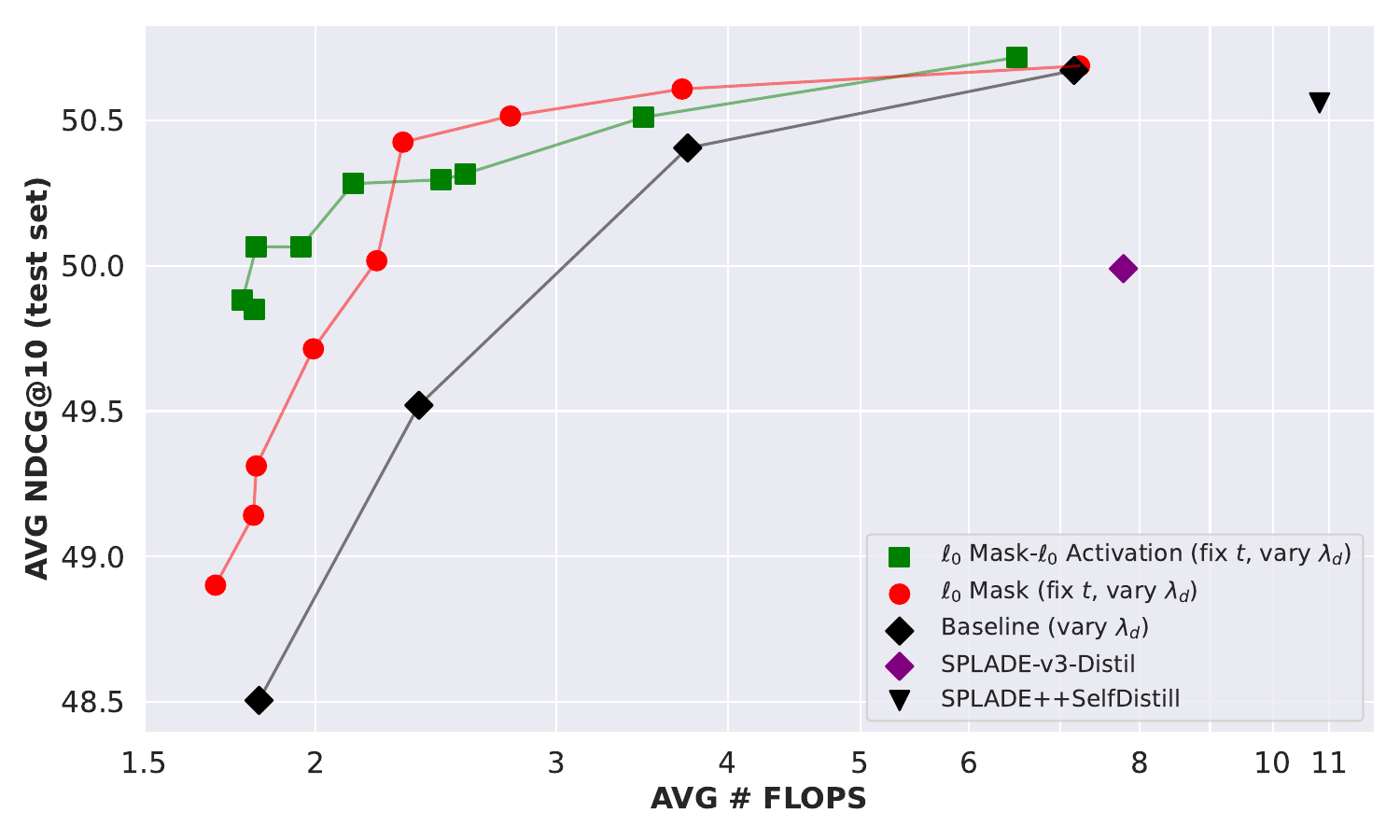}
    \vspace{-0.7cm}
    \caption{Search relevance vs efficiency, varying $\lambda_d$.}
    \label{fig:vary_lambda__flops_vs_ndcg}
    \vspace{-0.3cm}
\end{figure}

\begin{figure}[h]
    \centering
    \vspace{-0.3cm}
    \includegraphics[width=1\columnwidth]{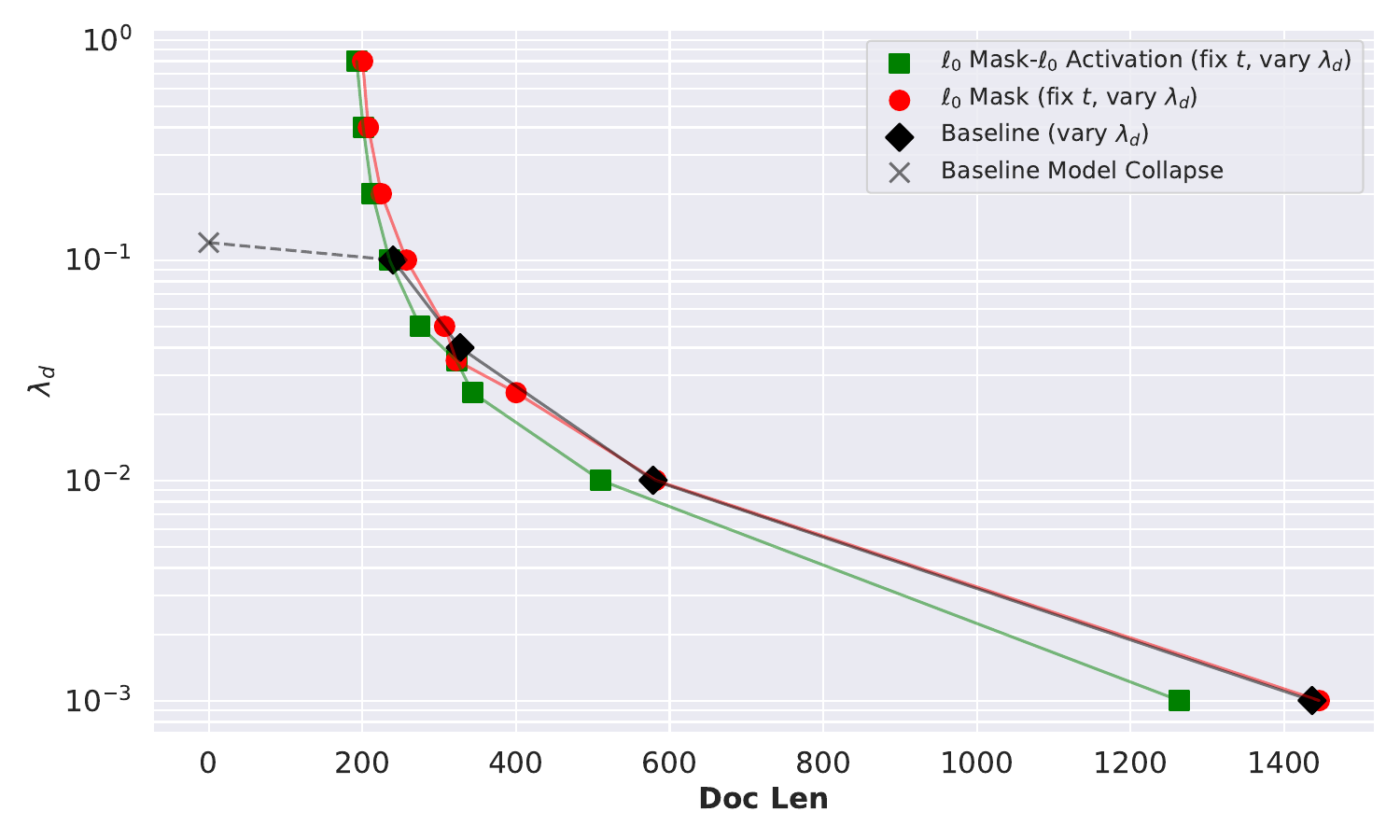}
    \vspace{-0.7cm}
    \caption{$\lambda_d$ vs encoded document sparsity. "x" denotes that baseline model collapses during training at $\lambda_d=0.12$.}
    \label{fig:vary_lambda__doc_len_vs_lambda}
    \vspace{-0.3cm}
\end{figure}

In Figures ~\ref{fig:vary_lambda__flops_vs_ndcg} and ~\ref{fig:vary_lambda__doc_len_vs_lambda}, we include representative models as baselines and vary hyperparameter $\lambda_d$ for comparison, with threshold $t$ at 200. The results in Figure ~\ref{fig:vary_lambda__flops_vs_ndcg} show that our proposed methods outperform the baselines in both performance and efficiency. For both $\ell_0$ activation and $\ell_0$ mask individually, the models achieve greater efficiency at the same performance level. Moreover, the results demonstrate that when applying $\ell_0$ activation with $\ell_0$ mask together, performance degrades more slowly as efficiency increases.

In Figure ~\ref{fig:vary_lambda__doc_len_vs_lambda}, we demonstrate how $\lambda_d$ affects the resulting Doc\_Len. We observed that the encoded Doc\_Len of the baseline model collapses to near zero during training when $\lambda_d$ increases beyond a certain level (0.12 in this case, denoted as "x" in the figure, with other settings fixed), as mentioned in our motivation in Section \ref{l0mask}. In contrast, our proposed methods remain stable as $\lambda_d$ increases, making model tuning and selection more flexible. This stability is achieved because the $\ell_0$ mask allows encoded documents with lengths below the threshold to avoid further penalty. From the side of Doc\_Len, applying $\ell_0$ activation helps model achieved encoded document sparsity.

\begin{figure}[h]
    \centering
    \includegraphics[width=1\columnwidth]{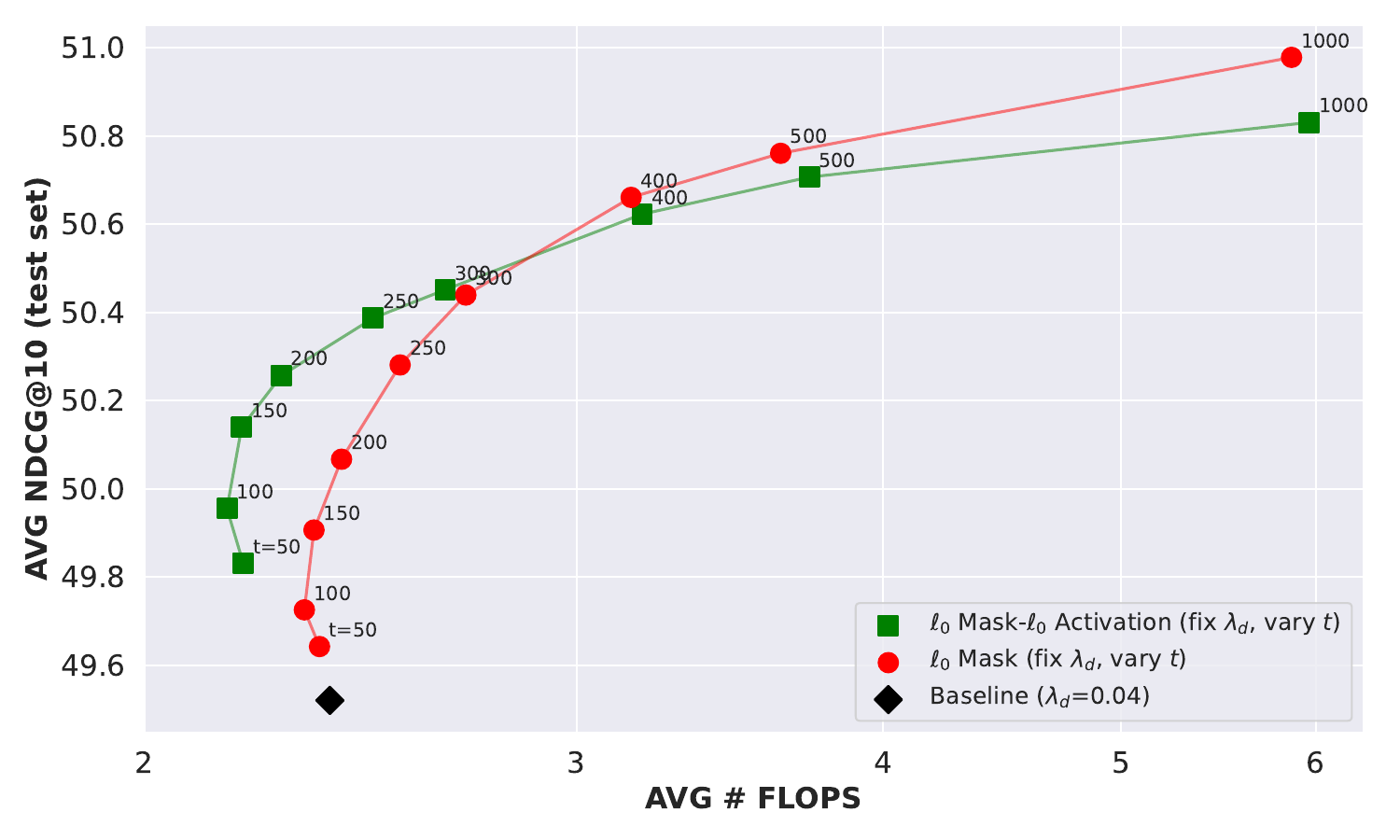}
    \vspace{-0.7cm}
    \caption{Search relevance vs sparsity, varying $t$.}
    \label{fig:vary_t__flops_vs_ndcg}
    \vspace{-0.3cm}
\end{figure}

\subsubsection{Varying Masked Threshold}
Furthermore, we discuss the effects of varying the threshold $t$ in our methods, with $\lambda_d$ fixed at 0.04. In Figure \ref{fig:vary_t__flops_vs_ndcg}, we observe that by varying $t$ in [50, 100, 150, 200, 250, 300, 400 500, 1000], our methods' performance increases when the threshold allows longer documents to avoid penalization. When the threshold approaches zero, the $\ell_0$ Mask degenerates to near baseline performance, where FLOPS directly affects all documents. Compared to Figure \ref{fig:vary_lambda__flops_vs_ndcg}, we can also find that at comparable FLOPS, the results obtained by varying $t$ are higher than those obtained by varying $\lambda_d$, indicating more exploration space and flexible model tuning.

\subsubsection{Varying $\ell_0$ Activation}
We investigate the impact of applying $\ell_0$ Activation multiple times (from 1 to 3) while keeping other settings fixed. Additionally, We explore two decoupling variants: de-$\ell_0$ activation that only uses $\ell_0$ activation for FLOPS calculation but remains unchanged for ranking objective, and inv-de-$\ell_0$ activation that only uses $\ell_0$ activation for ranking objective but remains unchanged for FLOPS calculation. Table \ref{tab:multi_log} shows the metrics for different numbers of $\ell_0$ Activation applications and the variant. We find that applying $\ell_0$ Activation decreases both FLOPS and Doc\_Len, indicating improved efficiency. However, applying it more than once begins to harm performance, which aligns with our motivation and analysis in Section \ref{l0activation}. Meanwhile, the decoupled versions show similar or lower performance but larger FLOPS and Doc\_Len, suggesting we should apply $\ell_0$ activation on both ranking objective and FLOPS.

\begin{table}[h]
\vspace{-0.2cm}
\caption{Comparison of applying variants of $\ell_0$ Activation.}
\centering
\label{tab:multi_log}
\vspace{-0.3cm}
\adjustbox{max width=0.48\textwidth}{
\begin{tabular}{cccc}
\hline
Model ($\lambda_d=0.035)$                  & NCDG@10 & FLOPS & Doc\_Len \\ \hline
Baseline & 49.52   & 2.39  & 327.23   \\
$\ell_0$ Activation                &  49.97       &  2.30     &  295.79       \\
de-$\ell_0$ Activation                &  49.77       &  2.52     &  345.63       \\
inv-de-$\ell_0$ Activation                &  49.54       &  2.36     &  326.58       \\
2-$\ell_0$ Activation         & 49.51   & 2.12  & 258.43   \\ 
3-$\ell_0$ Activation  & 48.83   & 2.05  & 245.64  \\ 
\hline
$\ell_0$ mask + $\ell_0$ Activation & 50.28   & 2.13  & 275.02  \\
$\ell_0$ mask + de-$\ell_0$ Activation                &  50.27      &  2.61     &  370.67       \\
$\ell_0$ mask + inv-de-$\ell_0$ Activation                &  49.34      &  2.18     &  283.78       \\
\hline
\end{tabular}
}
\vspace{-0.3cm}
\end{table}

\section{Conclusion}
In this paper, we propose simple yet effective $\ell_0$-inspired methods that achieve comparable performance to state-of-the-art sparse retrieval models in the inference-free setting, using same and fair training data and evaluation protocols. Moreover, we analyze the trade-off between retrieval effectiveness and computational efficiency, providing practical insights for implementation.

\appendix

\newpage
\bibliographystyle{ACM-Reference-Format}
\balance
\bibliography{sample-base}

\end{document}